\documentstyle{article}

\begin{document}

\begin{center}

{\large\bf PRECESSION OF A PARTICLE WITH ANOMALOUS MAGNETIC
MOMENT IN ELECTROMAGNETIC AND GRAVITATIONAL PP-WAVE FIELDS}

\medskip
{\bf
A. Balakin\footnotemark[1], V. Kurbanova\footnotemark[1]
and W. Zimdahl\footnotemark[2]}

\medskip
{\it\small\it
$^1$ Department of General Relativity and Gravitation, Kazan
University, Kremlevskaya str. 18, Kazan 420008, Russia

$^2$ Fachbereich Physik, Universit\"at Konstanz, PF M678, D-78457
Konstanz, Germany}

\end{center}

\medskip
{\noindent\small
We present an exact solution of the Bargmann-Michel-Telegdi (BMT)
equations for the dynamics of a spin particle in external
electromagnetic and gravitational pp-wave fields. We demonstrate
that an anomalous magnetic moment gives rise to an additional
spin rotation which is modulated both by the electromagnetic and
by the gravitational wave periodicities.}

\footnotetext[1]{\footnotesize e-mail: dulkyn@mail.ru}
\footnotetext[2]{\footnotesize e-mail: zimdahl@spock.physik.uni-konstanz.de}

\section{Introduction}

In 1959 Bargmann, Michel and Telegdi obtained the covariant evolution equations
for classical spin particles with anomalous magnetic moment~\cite{Bargmann}.
They described the precession of a spin four-vector in the framework of
Special Relativity as a generalization of the non-relativistic theory
by ~Thomas~\cite{Thomas} and Bloch~\cite{Bloch}.
Being completed by the spin-curvature coupling terms,
intro\-duced by Papapetrou~\cite{Papa}, the BMT model has also become a
starting point for numerous investigations of spin-particle dynamics in
General Relativity (see, e.g.,~\cite{Khrip} and references therein).

According to the BMT model, the dynamics of the spin vector is coupled to
the external electromagnetic (EM) fields and to
the four-velocity of the particle. In contrast, the evolution of the particle
four-velocity depends on the
fields only and is not influenced by the spin behaviour.
Consequently, in addition to the direct action of the external fields on
the spin, there exists an indirect influence via the particle dynamics.
Moreover, in a general relativistic context the electromagnetic field may
also provide the spacetime which is the arena for the covariant particle
dynamics.

While the original BMT equations have been derived for spatially homogeneus
EM fields, they are expected to be useful also for inhomogeneous situations
in which the corresponding gradients are sufficiently small
and the relevant effects are of first order in the spin variable~\cite{Khrip}.
The aim of this note is to study the cooperative effect of the
electromagnetic and gravitational wave (GW) fields on the spin rotation.
Assuming a pp-wave symmetry we find an exact analytic solution of the BMT
model without the weak field
approximation for the GW. For a particle with anomalous magnetic moment
we shall obtain a precession of the spin vector the frequency of which is
double-periodic, i.e., it inherits both the EM and GW field
periodicities.

\section{Basic Theory}

Our approach is based on the coupled set of Einstein's equations
for the gravitational field, Maxwell's equati\-ons for electromagnetic
field and the BMT equations for a charged spin particle with anomalous
magnetic moment. The particle will be regarded as a test particle, i.e.,
the EM and GW fields are background fields which are not influenced by the
particle motion.

\subsection{External pp-wave fields}

The external fields are assumed to have pp-wave symmetry.
The first one is a GW field with the metric~\cite{Misner}
\begin{equation}
ds^{2} = 2dudv - L^{2} \{2 \sinh{2\gamma} dx^2 dx^3 + \cosh2\gamma \left[{\rm e}^{2 \beta}(dx^2)^2 + {\rm e}^{-2\beta}
(dx^3)^2 \right]\}\ ,
\label{(1)}
\end{equation}
where
\begin{equation}
u = \frac{1}{\sqrt{2}}(ct - x^1)  \quad {\rm and}\quad v = \frac{1}{\sqrt{2}}(ct + x^1)
\label{(2)}
\end{equation}
are the retarded and the advanced times, respectively.
This gravitational field possesses a space-time symmetry
corresponding to the five-parametric isometry group $G_5$.
Moreover, the three Killing vectors
\begin{equation}
\xi_{(v)}^i = \delta_v^i \ , \quad \xi_{(2)}^i = \delta_2^i \ ,
\quad \xi_{(3)}^i = \delta_3^i \ ,
\label{(3)}
\end{equation}
form an Abelian subgroup of $G_5$ ~\cite{Kramer}.

The second external field is an electromagnetic field with the Maxwell tensor
\begin{equation}
F_{jk} =  \dot{A}_2(u)\left(\delta^{u}_{j} \delta^{2}_{k}
- \delta^{2}_{j}\delta^{u}_{k} \right) +
\dot{A}_3(u) \left(\delta^{u}_{j} \delta^{3}_{k}
- \delta^{3}_{j}\delta^{u}_{k} \right) \ ,
\label{(8)}
\end{equation}
where a dot denotes the derivative with respect to retarded time $u$.
The functions $\beta(u)$ and $\gamma(u)$ in~(\ref{(1)})
and $A_2(u)$ and $A_3(u)$ in (\ref{(8)}) are arbitrary.
In particular, they may be periodic functions of the retarded time.
The field strength tensor~(\ref{(8)}) satisfies Maxwell's  equations
in a pp-wave gravitational back\-ground identi\-cally~\cite{Sibgatullin}.
It describes a null field, i.e. the first and the second invariants are
equal to zero, and it inherits the symmetry of the GW background.
Alterna\-tively, the five functions, $L(u)$, $\beta(u)$, $\gamma(u)$,
$A_2(u)$ and $A_3(u)$ may also
be regarded as solutions of the Einstein-Maxwell
equations~\cite{Misner,Kramer,Sibgatullin}:
\begin{equation}
\ddot{L}
+ L \left(\dot{\beta}^2 \cosh^2{2\gamma} + \dot{\gamma}^2 \right) =
\frac{G}{2c^2} \left( g^{\alpha \beta} \dot{A}_{\alpha} \dot{A}_{\beta}\right) \ .
\label{(4)}
\end{equation}
Greek indices run from 2 to 3. An explicit analyti\-cal solutions of this equation
was found by Brdi\v{c}ka (\cite{Brd}, cf.~\cite{Kramer}).
Some other solutions are mentioned in~\cite{Zakharov}.
However, it will not be necessary to specify the functions
$\beta(u), \gamma(u)$ and $A_{\alpha}(u)$ in the following.

\subsection{BMT equations}

The BMT model relies on the set of equations:
\begin{equation}
\frac{D{U^{i}}}{D\tau} = \frac{e}{mc^2} F^i_{ \cdot k} \ U^{k} \ ,
\label{(9)}
\end{equation}
\begin{equation}
\frac{D{S^{i}}}{D\tau} = \frac{e}{mc^2} \left[ \ \frac{g}{2} F^i_{ \cdot k} \
S^{k} + \left(\frac{g}{2} - 1 \right) U^i F_{kl} S^k U^l \right] \ .
\label{(10)}
\end{equation}
Here, $U^i$ is the particle four-velocity vector,
$S^i$ is the spin four-vector, dual to the spin-tensor $S_{lm}$,
\begin{equation}\label{spin}
S^i\equiv\frac 12\epsilon^{iklm}U_k S_{lm} \ .
\end{equation}
The symbol $D$ denotes the covariant differential,
$e$ is the charge of the particle,
$m$ is its mass,
$c$ is the velocity of light and
$g$ is the gyromagnetic ratio.
An anomalous magnetic moment is characterized by
$\frac g2 \neq 1$.

Equation~(\ref{(9)}) is the standard equation of motion for a charged particle
in a Lorentz force field. The particle motion does not depend on the spin.
Equation~(\ref{(10)}) describes the spin precession.
Equations~(\ref{(9)}) and~(\ref{(10)}) imply that $U^i S_i$ is constant.
Orthogonality of the spin and the four-velocity
requires  $U^i S_i=0$.
We will start our analysis by integrating~(\ref{(9)}) since this equation is decoupled.

\section{Four-velocity vector}

Since the Lorentz force
is orthogonal to particle four-velocity,
the first quadratic integral of the system~(\ref{(9)}) is
\begin{equation}
g^{ik} U_i U_k =  {\rm const} \equiv 1 \ .
\label{(11)}
\end{equation}
This relation may be solved for the $U_u$ component
of the particle four-velocity,
\begin{equation}
U_u = \frac{1}{2U_v} \left[ 1 - g^{\alpha \sigma}(u) \ U_\alpha U_\sigma\right]\ ,
\label{(12)}
\end{equation}
where $\alpha ,\sigma\ = 2, 3$.
Since the GW field is characterized by a
covariantly constant null Killing vector $\xi^i_{(v)} = \delta^i_v$,
and since $F_{ik}\xi^i_{(v)} = 0$,
the contraction of the four-velocity $U_i$
with $\xi^i_{(v)}$ is an
integral of motion (see e.g.~\cite{Misner}):
\begin{equation}
\xi^i_{(v)} U_i = U_v = C_v = {\rm const}\ .
\label{(13)}
\end{equation}
With the help of
\begin{equation}
\frac{d u}{d\tau} = U^u = U_v = C_v \ ,
\label{(14)}
\end{equation}
we get the linear relation
\begin{equation}
\tau = \tau_0 + \frac{u}{C_v}
\qquad\ (C_v \neq 0)\ ,
\label{(15)}
\end{equation}
which may be used to reparametrize the remaining equations.
For the transverse components of the four-velocity vector
we have
\begin{equation}
\frac{dU_{\alpha}}{d u}=\frac{e}{mc^2} F_{\alpha u}\ .
\label{(46)}
\end{equation}
Using the Maxwell tensor~(\ref{(8)}), we obtain immediately
\begin{equation}
U_\alpha(u) = C_\alpha - \frac{e}{mc^2} A_\alpha(u)\ .
\label{(47)}
\end{equation}
The transverse components of the four-velocity depend on $u$ via the vector potential.
In particular, for an oscillating field these components oscillate with the EM wave frequency.

\section{Spin four-vector}

As already mentioned, the equations ~(\ref{(9)}) and~(\ref{(10)})
admit $U_i S^i=0$ as an integral of motion.
This can be used
to express the $u$-component of the spin four-covector $S_i$
in terms of $S_v$, $S_\alpha$ and $U^i$:
\begin{equation}\label{S_u}
S_u = -\frac 1{C_v}\left[ g^{\alpha\sigma} U_\alpha S_\sigma + U_u S_v\right]\ .
\end{equation}
A second integral of motion is
$S_i S^i = {\rm const}$~\cite{Bargmann}.

\subsection{General case}

With the reparametrization~(\ref{(15)})
the  equations for
$S_\alpha$ and $S_v$ are
\begin{equation}
\dot{S}_{\alpha} =
\frac 12 g^{\sigma\rho}\dot{g}_{\rho\alpha} \left(S_\sigma
- U_\sigma \frac{S_v}{C_v}\right)
+ \frac{e}{mc^2 C_v} \left[ \ \frac{g}{2} F_{\alpha k} \ S^{k}
\right.
+ \left.\left(\frac g2 - 1 \right) U_\alpha F_{kl} S^k U^l \right]\ ,
\label{(51)}
\end{equation}
and
\begin{equation}
\dot{S}_v =
\left(\frac{g}{2} - 1 \right) \frac{e}{mc^2} F_{kl} S^k U^l \ ,
\label{(52)}
\end{equation}
respectively.
After the substitution
\begin{equation}
S_\alpha = \frac{1}{C_v} \left( X_\alpha + S_v U_\alpha \right)\ ,
\label{(54)}
\end{equation}
by which we separate the degrees of freedom which are proportional to the
four-velocity,
we obtain
\begin{equation}
\dot{X}_{\alpha} =
\frac{1}{2} g^{\sigma\rho} \dot{g}_{\rho\alpha} X_\sigma
- \chi S_v \dot{A}_{\alpha} \ ,
\label{(55)}
\end{equation}
and
\begin{equation}
\dot{S}_v = -
\chi \dot{A}_{\alpha} g^{\alpha \beta} X_{\beta} \
\label{(521)}
\end{equation}
for the new set of variables.
Here we have introduced the constant
\begin{equation}
\chi \equiv \frac{e}{mc^2} \left(\frac{g}{2} - 1 \right) \ .
\label{(614)}
\end{equation}
Equation~(\ref{(55)}) may be written in the matrix form
\begin{equation}
\dot{{\bf X}} = {\bf B} \ {\bf X} - \chi S_v \dot{{\bf A}}\ ,
\label{(56)}
\end{equation}
where ${\bf X}$ and $\dot{{\bf A}}$ are column vectors and ${\bf B}$ is
a two-dimensional matrix with the elements
\begin{eqnarray}\label{B-coef}
B_2^2 &=& \frac{\dot{L}}L + \cosh^2(2\gamma )\dot{\beta}\ , \nonumber\\
B_3^3 &=& \frac{\dot{L}}L - \cosh^2(2\gamma )\dot{\beta}\ , \nonumber\\
B_2^3 &=& {\rm e}^{ 2\beta}\left[\dot{\gamma} - \sinh(2\gamma)\cosh(2\gamma)\dot{\beta}\right]\ ,  \nonumber\\
B_3^2 &=& {\rm e}^{-2\beta}\left[\dot{\gamma} + \sinh(2\gamma)\cosh(2\gamma)\dot{\beta}\right]\ .
\end{eqnarray}
In order to simplify the system~(\ref{(56)}) we introduce the vector ${\bf Y}$ by
\begin{equation}
{\bf X} =  {\bf T} \cdot {\bf Y}\ ,
\label{(58)}
\end{equation}
where the two-dimensional matrix ${\bf T}$ is required to satisfy
the differential equation
\begin{equation}
\dot{{\bf T}} = {\bf B} \cdot {\bf T}\ .
\label{(59)}
\end{equation}
The physical background for the ansatz (\ref{(58)}) is the expectation that the rotational part
${\bf X}$ itself may be decomposed into two different types of motion.
The resulting equation for the
new unknown column vector ${\bf Y}$ is
\begin{equation}
\dot{{\bf Y}} =  - \chi S_v {\bf T}^{-1}\dot{{\bf A}}  \ .
\label{(60)}
\end{equation}
Guided by previous experience for similar problems
(\cite{Balakin,BK_Volga98,BKFP}), we find that
\begin{equation}
{\bf T}=
L \cdot \left(\begin{array}{cc}e^{\beta } & 0 \\ 0 & e^{-
\beta}\end{array}\right) \cdot
\left(\begin{array}{cc}
\cosh \gamma  & \sinh \gamma  \\
\sinh \gamma  & \cosh \gamma
\end{array}\right)\times
\left(\begin{array}{cc}
\cos \psi  & -\sin \psi  \\
\sin \psi  & \cos \psi
\end{array} \right)\ ,
\label{(61)}
\end{equation}
where
\begin{equation}
\psi \equiv \int\limits_0^u \dot{\beta} \sinh 2\gamma du \ .
\label{(62)}
\end{equation}
One can check by direct calculation that the matrix~(\ref{(61)}) indeed
satisfies the equation~(\ref{(59)}).
The structure of the third matrix on the right-hand side of (\ref{(61)}) suggests the interpretation as
a gravitationally induced rotation with a phase $\psi(u)$
and a frequency $\dot{\psi}(u)$. This motion is independent of the magnetic moment.
The determi\-nants of each of the three two-dimensional matrices in~(\ref{(61)})
are equal to one. In the absence of the GW field all of them are identical to the unity matrix and
we have
\begin{equation}
{\bf T}(0) =  {\bf I} \equiv
\left(\begin{array}{cc}
1  &  0  \\
0  &  1
\end{array} \right) \ .
\label{(62a)}
\end{equation}
For the polarization $\gamma  =0$ the matrix ${\bf T}$
in (\ref{(61)}) reduces to
\begin{equation}
{\bf T}(u) =
L \cdot \left(\begin{array}{cc}e^{\beta } & 0 \\ 0 & e^{-
\beta}\end{array}\right) \qquad \qquad
(\gamma =0)\ .
\label{(62b)}
\end{equation}
A corresponding simplification is obtained for the polari\-zation $\beta =0$.
For polarized GWs ($\gamma=0$ or $\beta=0$) the quantity $\psi$ in
(\ref{(62)}) vanishes.

In a next step we use ${\bf T}$ as given in (\ref{(61)}) to integrate the
equations for $S_v$, $Y_2$ and $Y_2$.
After the substitu\-ti\-on~(\ref{(58)}) the system (\ref{(52)}) and (\ref{(60)}) becomes
\begin{equation}
\dot{S}_v = - \chi \dot{A}_{\alpha} g^{\alpha \beta}T_\beta^\delta \cdot
Y_\delta \ ,
\label{(611)}
\end{equation}
\begin{equation}
\dot{Y}_\rho = - \chi (T^{-1})_\rho^\alpha \dot{A}_{\alpha}\cdot S_v \ .
\label{(612)}
\end{equation}
Introducing the new coefficients
\begin{equation}
\begin{array}{c}
b_2\equiv \chi L^{-1}
[\dot{A}_2 {\rm e}^{-\beta}(\sin\psi \sinh\gamma - \cos\psi \cosh\gamma) \\
\qquad + \dot{A}_3 {\rm e}^\beta ( \cos\psi \sinh\gamma - \sin\psi\cosh\gamma)]\ ,
\end{array}
\label{(615)}
\end{equation}
and
\begin{equation}
\begin{array}{c}
b_3\equiv\chi L^{-1}
[\dot{A}_2{\rm e}^{-\beta}( \cos\psi \sinh\gamma + \sin\psi \cosh\gamma)  \\
\qquad + \dot{A}_3{\rm e}^\beta (-\cos\psi \cosh\gamma - \sin\psi\sinh\gamma)]\ ,
\end{array}
\label{(616)}
\end{equation}
the equations (\ref{(611)}) and (\ref{(612)}) simplify to
\begin{eqnarray}
\dot{S}_v &=& -(b_2 Y_2 + b_3 Y_3)\ , \nonumber\\
\dot{Y}_2 &=& b_2 S_v \ , \nonumber\\
\dot{Y}_3 &=& b_3 S_v \ .
\label{(618)}
\end{eqnarray}
It is straightforward to check that there exists a quadra\-tic integral
\begin{equation}
Y_2^2 + Y_3^2 + S_v^2 = {\rm const} \ .
\label{(619)}
\end{equation}
While for $\chi = 0$, equivalent to $g=2$, one has $b_2=b_3=0$ and
the quantities $S_v$, $Y_2$ and $Y_2$ are separately conserved, an anomalous magnetic moment gives rise to a rotation which preserves only the combination (\ref{(619)}).

This completes our general analysis of the BMT equations.
We have decomposed the entire spin dy\-namics into a part proportional
to the four-velocity (cf.~(\ref{(54)}) with~(\ref{(47)})) and two rotations.
The first rotation is described by the matrix ${\bf T}$.
It is purely gravitational and does not depend on the EM field.
In particular, it is independent of the magnetic momentum and it vanishes for polarized GWs.
The second rotation is characterized by the equations (\ref{(615)})-(\ref{(619)}),
which are the main results of the present paper.
This motion is a result of the coupling of the anomalous magnetic moment to the EM field which is modulated
by the GW field.
In the absence of the GW and for
$\chi=0$ the additional rotational degrees of freedom disappear and the
dynamics reduces to (cf.~(\ref{(54)}) with~(\ref{(47)}))
\begin{equation}
S_\alpha(u) - S_\alpha(0) = - \frac{eS_v(0)}{mc^2 C_v} A_{\alpha}(u) \ ,
\label{(543)}
\end{equation}
i.e., the spin inherits the $u$-dependence of the electro\-magnetic field.
This type of motion is similar to the standard spin precession in an EM field,
but there is an additional modulation with the frequency of the field.
To obtain further insight, we will consider a two-dimensional exam\-ple in the following subsection.

\subsection{Two-dimensional motion}

This model is based on three simplifications.
Firstly, the GW is supposed to be polarized according to $g_{23} = 0$, i.e., $\gamma = 0$.
Secondly, for the electromagnetic field we assume a polarization $A_3 = 0$.
Thirdly, we simplify the initial data such that $U_3(0) =0$ and $S_3(0) = 0$.
These requirements guarantee that the components $U_3(u)$ and $S_3(u)$ remain zero during the particle evolution
and therefore $X_3(u) = Y_3(u) \equiv 0 $.
In this case the substitutions~(\ref{(54)}) and~(\ref{(58)}) correspond to
\begin{equation}
Y_2(u) \equiv \frac{1}{L} e^{-\beta}(C_v S_2 - U_2 S_v ) \ .
\label{(73)}
\end{equation}
The coupled variables $Y_2(u)$ and $S_v(u)$
satisfy the equations
\begin{equation}
\dot{Y}_2 = - \Omega_A(u) S_v \  \quad {\rm and} \quad
\dot{S}_v = \Omega_A(u) Y_2 \ ,
\label{(71)}
\end{equation}
where $\Omega_A(u)$ is given by
\begin{equation}
\Omega_A(u) \equiv  \chi \dot{A}_2 \ e^{-\beta} \ .
\label{(75)}
\end{equation}
The solution of the system~(\ref{(71)}) represents a two-dimen\-sional rotation for which we explicitly obtain
\begin{eqnarray}
S_v(u) &=& (C_v E_2 - C_2 E_v) \sin{\Phi} +  E_v \cos{\Phi} \ ,
\label{(76)}\\
S_2(u) &=& E_2 \left[ \left( C_2 - \frac{e}{mc^2} A_2 \right) \sin{\Phi} +
L e^{\beta} \cos{\Phi} \right] \nonumber\\
&&+\frac{E_v}{C_v} \left( C_2 - \frac{e}{mc^2} A_2 \right)(\cos{\Phi} - C_2\sin{\Phi})\nonumber\\
&&- \frac{E_v}{C_v}\cdot L e^{\beta} [C_2 \cos{\Phi} + \sin{\Phi}]\ ,
\label{(77)}
\end{eqnarray}
where
$\dot{\Phi} \equiv  \Omega_A(u)$. The quantity $\Omega_A(u)$ plays the role of a time dependent frequency
for the anomalous spin precession. If the vector potential $A_2$ and the GW parameter $\beta$ are periodic,
$\Omega_A$ will be double-periodic.

\section{Conclusions}

External electromagnetic and gravitational fields induce an additional precession of a classical
spin particle if the latter has an anomalous magnetic moment.
This was explicitly demonstrated by solving the BMT equati\-ons~(\ref{(9)}) and~(\ref{(10)}) in the pp-wave fields
(\ref{(1)}) and (\ref{(8)}) without resorting to a weak field approximation.
The correspon\-ding anomalous precession frequency (\ref{(75)})
is the result of a gravitationally modulated coupling of the anomalous magnetic momentum to the electromagnetic field.
It is is periodic with the periods of the fields.

\subsection*{Acknowledgement}

This work was supported by the Deutsche Forschungs\-gemeinschaft and by NATO.

\end{document}